\documentclass{ws-ijmpa}

\begin{document}

\markboth{Zhi-zhong Xing} {TeV Neutrino Physics at the Large
Hadron Collider}

%
\catchline{}{}{}{}{}
%

\title{TEV NEUTRINO PHYSICS AT THE LARGE HADRON COLLIDER\footnote{Invited
talk given at the International Conference on Particle Physics,
Astrophysics and Quantum Field Theory: 75 Years since Solvay,
27---29 November 2008, Singapore.}}

\author{ZHI-ZHONG XING}

\address{Institute of High Energy Physics
and Theoretical Physics Center for Science Facilities, \\ Chinese
Academy of Sciences, P.O. Box 918, Beijing 100049, China \\
xingzz@ihep.ac.cn}

\maketitle


\begin{abstract}
I argue that TeV neutrino physics might become an exciting frontier
of particle physics in the era of the Large Hadron Collider (LHC).
The origin of non-zero but tiny masses of three known neutrinos is
probably related to the existence of some heavy degrees of freedom,
such as heavy Majorana neutrinos or heavy Higgs bosons, via a
TeV-scale seesaw mechanism. I take a few examples to illustrate how
to get a balance between theoretical naturalness and experimental
testability of TeV seesaws. Besides possible collider signatures at
the LHC, new and non-unitary CP-violating effects are also expected
to show up in neutrino oscillations for type-I, type-(I+II) and
type-III seesaws at the TeV scale.

\keywords{TeV seesaws; Collider signatures; Violation of
unitarity; Neutrino oscillations.}
\end{abstract}


\section{Why TeV Seesaws?}

Enrico Fermi elaborated a coherent theory of the beta decay and
published it in {\it La Ricerca Scientifica} in December
1933,\cite{Fermi} just two months after the Solvay Congress in
October 1933. In this seminal paper, Fermi postulated the existence
of a new force for the beta decay by combining three brand-new
concepts
--- Pauli's neutrino hypothesis, Dirac's idea about the creation of
particles, and Heisenberg's idea that the neutron was related to the
proton. Today, we have achieved a standard theory of electroweak
interactions at the Fermi scale ($\sim 100$ GeV), although it is
unable to tell us much about the intrinsic physics of electroweak
symmetry breaking and the origin of non-zero but tiny neutrino
masses. We are expecting that the LHC will soon bring about a
revolution in particle physics at the TeV scale ($\sim 1000$ GeV)
--- a new energy frontier that we humans have never reached before
within a laboratory. Can the LHC help solve the puzzle of neutrino
mass generation? We do not yet know the answer to this question. But
let us hope so. I personally foresee that {\it TeV neutrino physics}
might become an exciting direction in the era of the LHC.

Among many theoretical and phenomenological ideas towards
understanding why the masses of three known neutrinos are so
small,\cite{PDG} the seesaw picture seems to be most natural and
elegant. Its key point is to ascribe the smallness of neutrino
masses to the existence of some new degrees of freedom heavier than
the Fermi scale, such as heavy Majorana neutrinos or heavy Higgs
bosons. Three typical seesaw mechanisms are illustrated in Fig. 1,
and some other variations or combinations are possible. The energy
scale where a seesaw mechanism works is crucial, because it is
relevant to whether this mechanism is theoretically natural and
experimentally testable. Between Fermi and Planck scales, there
might exist two other fundamental scales: one is the scale of a
grand unified theory (GUT) at which strong, weak and electromagnetic
forces can be unified, and the other is the TeV scale at which the
unnatural gauge hierarchy problem of the standard model (SM) can be
solved or at least softened by new physics. Many theorists argue
that the conventional seesaw scenarios are natural because their
scales (i.e., the masses of heavy degrees of freedom) are close to
the GUT scale. If the TeV scale is really a fundamental scale, may
we argue that the TeV seesaws are natural? In other words, we are
reasonably motivated to speculate that possible new physics existing
at the TeV scale and responsible for the electroweak symmetry
breaking might also be responsible for the origin of neutrino
masses.\cite{ICHEP08} It is interesting and meaningful in this sense
to investigate and balance the ``naturalness" and ``testability" of
TeV seesaws at the energy frontier set by the LHC.
\begin{figure*}[t]
\centering
\includegraphics[width=128mm]{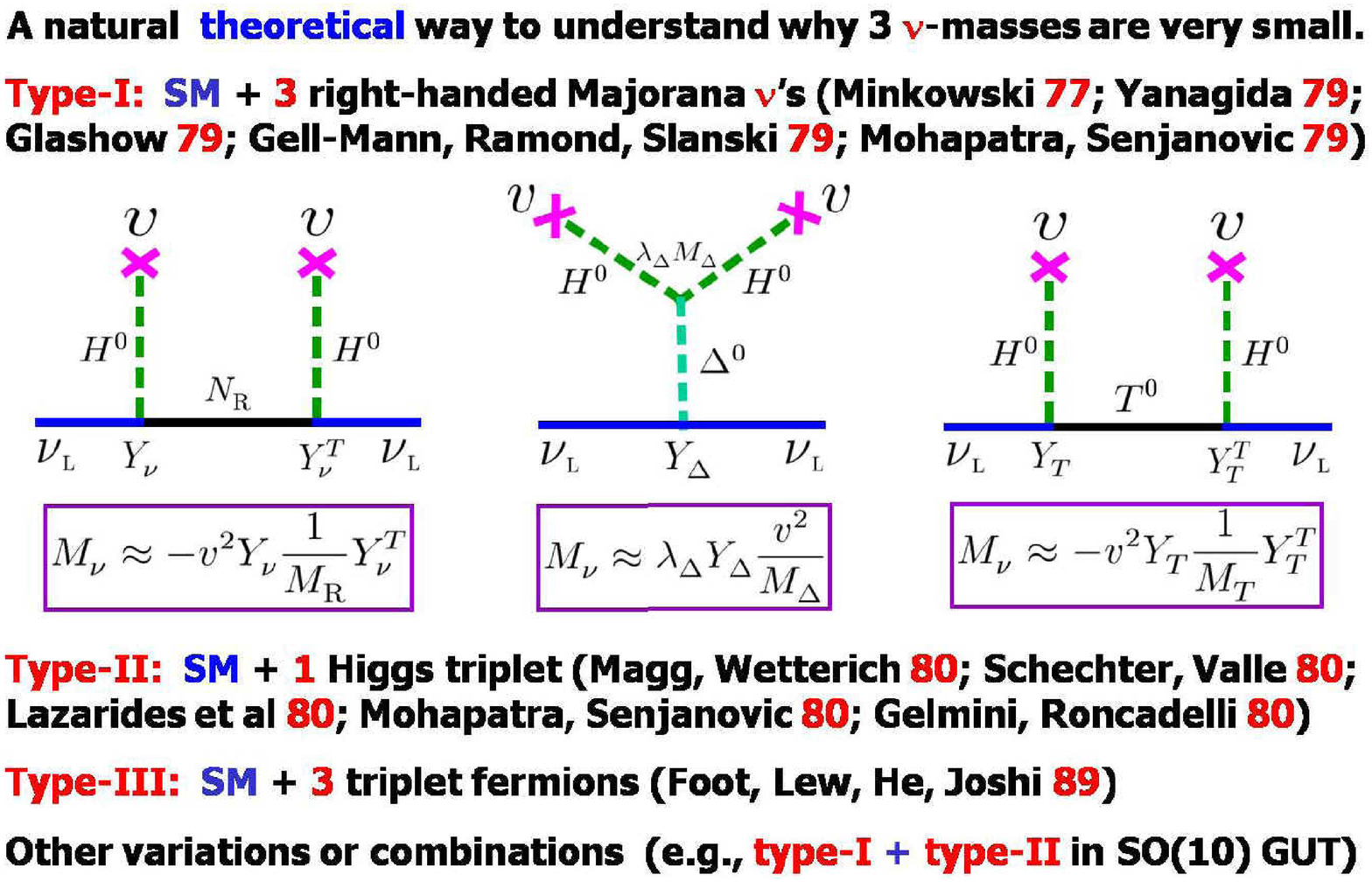}
\caption{Three types of seesaw mechanisms to understand non-zero but
tiny neutrino masses.}
\end{figure*}

\section{Naturalness and Testability}

As shown in Fig. 1, the type-I seesaw mechanism gives a natural
explanation of the smallness of neutrino masses by introducing three
heavy right-handed Majorana neutrinos, while the type-II seesaw
mechanism is to extend the SM by including one $SU(2)_L$ Higgs
triplet. One may in general combine the two mechanisms by assuming
the existence of both the Higgs triplet and right-handed Majorana
neutrinos, leading to a ``hybrid" seesaw scenario which will be
referred to as the type-(I+II) seesaw mechanism. The gauge-invariant
neutrino mass terms in such a type-(I+II) seesaw model can be
written as
\begin{eqnarray}
-{\cal L}^{}_{\rm mass} \; = \; \overline{l^{}_{\rm L}} Y^{}_\nu
\tilde{H} N^{}_{\rm R} + \frac{1}{2} \overline{N^{c}_{\rm R}}
M^{}_{\rm R} N^{}_{\rm R} + \frac{1}{2} \overline{l^{}_{\rm L}}
Y^{}_\Delta \Delta i\sigma^{}_2 l^c_{\rm L} + {\rm h.c.} \; ,
\end{eqnarray}
where $M^{}_{\rm R}$ is the mass matrix of right-handed Majorana
neutrinos, and
\begin{equation}
\Delta \; \equiv \; \left(\begin{matrix} H^- & - \sqrt{2} ~ H^0 \cr
\sqrt{2} ~ H^{--} & -H^- \end{matrix} \right)
\end{equation}
denotes the $SU(2)_L$ Higgs triplet. After spontaneous gauge
symmetry breaking, we obtain the neutrino mass matrices $M^{}_{\rm
D} = Y^{}_\nu v/\sqrt{2}$ and $M^{}_{\rm L} = Y^{}_\Delta
v^{}_\Delta$, where $\langle H \rangle \equiv v/\sqrt{2}$ and
$\langle \Delta \rangle \equiv v^{}_\Delta$ correspond to the vacuum
expectation values of the neutral components of $H$ and $\Delta$.
Then Eq. (1) can be rewritten as
\begin{eqnarray}
-{\cal L}^\prime_{\rm mass} \; = \; \frac{1}{2} ~ \overline{\left(
\nu^{}_{\rm L} ~N^c_{\rm R}\right)} ~ \left( \begin{matrix}
M^{}_{\rm L} & M^{}_{\rm D} \cr M^T_{\rm D} & M^{}_{\rm R}
\end{matrix}\right) \left( \begin{matrix} \nu^c_{\rm L} \cr N^{}_{\rm R}
\end{matrix} \right) + {\rm h.c.} \; .
\end{eqnarray}
The $6\times 6$ neutrino mass matrix in Eq. (3) is symmetric and can
be diagonalized by the following unitary transformation:
\begin{eqnarray}
\left( \begin{matrix} V & R \cr S & U \end{matrix} \right)^\dagger
\left( \begin{matrix} M^{}_{\rm L} & M^{}_{\rm D} \cr M^T_{\rm D} &
M^{}_{\rm R} \end{matrix}\right) \left(\begin{matrix} V & R \cr S &
U \end{matrix} \right)^*  = \left( \begin{matrix} \widehat{M}^{}_\nu
& {\bf 0} \cr {\bf 0} & \widehat{M}^{}_N \end{matrix}\right) \; ,
\end{eqnarray}
where $\widehat{M}^{}_\nu = {\rm Diag}\{m^{}_1, m^{}_2, m^{}_3\}$
with $m^{}_i$ being the masses of three light neutrinos $\nu^{}_i$
and $\widehat{M}^{}_N = {\rm Diag}\{M^{}_1, M^{}_2, M^{}_3\}$ with
$M^{}_i$ being the masses of three heavy neutrinos $N^{}_i$. Note
that $V^\dagger V + S^\dagger S = VV^\dagger + RR^\dagger = {\bf 1}$
holds as a consequence of the unitarity of this transformation.
Hence $V$, the flavor mixing matrix of three light neutrinos, must
be non-unitary if $R$ and $S$ are non-zero.

\subsection{Type-I seesaw}

The type-I seesaw scenario can be obtained from Eqs. (1)---(4) by
switching off the Higgs triplet. In this case, $M^{}_{\rm L} =
{\bf 0}$ and $R \sim S \sim M^{}_{\rm D}/M^{}_{\rm R}$ hold,
leading to the approximate seesaw formula
\begin{equation}
M^{}_\nu \; \equiv \; V \widehat{M}^{}_\nu V^T \; \approx \; -
M^{}_{\rm D} M^{-1}_{\rm R} M^T_{\rm D} \; .
\end{equation}
The deviation of $V$ from unitarity is measured by $RR^\dagger/2$
and has been neglected in this expression. Let us consider two
interesting possibilities:

(1) $M^{}_{\rm D} \sim {\cal O}(10^2)$ GeV and $M^{}_{\rm R} \sim
{\cal O}(10^{15})$ GeV to get $M^{}_\nu \sim {\cal O}(10^{-2})$ eV.
In this conventional and {\it natural} case, $R \sim S \sim {\cal
O}(10^{-13})$ holds. Hence the non-unitarity of $V$ is only at the
${\cal O}(10^{-26})$ level, too small to be observed.

(2) $M^{}_{\rm D} \sim {\cal O}(10^2)$ GeV and $M^{}_{\rm R} \sim
{\cal O}(10^{3})$ GeV to get $M^{}_\nu \sim {\cal O}(10^{-2})$ eV.
In this {\it unnatural} case, a significant ``structural
cancellation" has to be imposed on the textures of $M^{}_{\rm D}$
and $M^{}_{\rm R}$. Because of $R \sim S \sim {\cal O}(0.1)$, the
non-unitarity of $V$ can reach the percent level and may lead to
some observable effects.

Now let us discuss how to realize the above ``structural
cancellation" for the type-I seesaw mechanism at the TeV scale.
Taking the flavor basis of $M^{}_{\rm R} = \widehat{M}^{}_N$, one
may easily show that $M^{}_\nu$ in Eq. (5) vanishes if
\begin{equation}
M^{}_{\rm D} \; = \; m \left( \begin{matrix} y^{}_1 & y^{}_2 &
y^{}_3 \cr \alpha y^{}_1 & \alpha y^{}_2 & \alpha y^{}_3 \cr \beta
y^{}_1 & \beta y^{}_2 & \beta y^{}_3 \cr \end{matrix} \right)
~~~~~~~~~ {\rm and} ~~~~~~~~~~ \sum^3_{i=1} \frac{y^2_i}{M^{}_i} \;
= \; 0
\end{equation}
simultaneously hold.\cite{T1} Tiny neutrino masses can be generated
from tiny corrections to the texture of $M^{}_{\rm D}$ in Eq. (6).
For example, $M^\prime_{\rm D} = M^{}_{\rm D} - \epsilon X^{}_{\rm
D}$ with $M^{}_{\rm D}$ given above and $\epsilon$ being a small
dimensionless parameter (i.e., $|\epsilon| \ll 1$) will yield
\begin{equation}
M^\prime_\nu \; \approx \; -M^\prime_{\rm D} M^{-1}_{\rm R}
M^{\prime T}_{\rm D} \; \approx \; \epsilon \left( M^{}_{\rm D}
M^{-1}_{\rm R} X^T_{\rm D} + X^{}_{\rm D} M^{-1}_{\rm R} M^T_{\rm D}
\right) \; ,
\end{equation}
from which $M^\prime_\nu \sim {\cal O}(10^{-2})$ eV can be obtained
by adjusting the size of $\epsilon$. We learn the following lessons
from this simple exercise:
\begin{itemize}
\item     Two necessary conditions must be satisfied in order to
test a type-I seesaw model at the LHC: (a) $M^{}_i$ are of ${\cal
O}(1)$ TeV or smaller; and (b) the strength of light-heavy neutrino
mixing (i.e., $M^{}_{\rm D}/M^{}_{\rm R}$) are large enough.
Otherwise, it would be impossible to produce and detect $N^{}_i$ at
the LHC.

\item     The collider signatures of $N^{}_i$ are essentially
decoupled from the mass and mixing parameters of three light
neutrinos $\nu^{}_i$. For instance, the small parameter $\epsilon$
in Eq. (7) has nothing to do with the ratio $M^{}_{\rm
D}/M^{}_{\rm R}$.

\item     The non-unitarity of $V$ might lead to some
observable effects in neutrino oscillations and other
lepton-flavor/number-violating processes, provided $M^{}_{\rm
D}/M^{}_{\rm R} \lesssim {\cal O}(0.1)$ holds. More discussions will
be given later.

\item     The clean LHC signatures of heavy Majorana neutrinos are
the $\Delta L =2$ like-sign dilepton events,\cite{Senjanovic} such
as $pp \to W^{*\pm} W^{*\pm} \to \mu^\pm \mu^\pm jj$ (a collider
analogue to the neutrinoless double-beta decay) and $pp \to W^{*\pm}
\to \mu^\pm N_i \to \mu^\pm \mu^\pm jj$ (a dominant channel due to
the resonant production of $N^{}_i$).
\end{itemize}
Some naive numerical calculations of possible LHC events for a
single heavy Majorana neutrino have been done in the
literature,\cite{Han} but they only serve for illustration because
such a minimal version of the type-I seesaw scenario is actually
unrealistic.

\subsection{Type-II seesaw}

The type-II seesaw scenario can be obtained from Eqs. (1)---(4) by
switching off the right-handed Majorana neutrinos and taking account
of a simple potential of the Higgs doublet and triplet:
\begin{equation}
{\cal V} \; =\; -\mu^2 H^\dagger H + \lambda \left( H^\dagger H
\right)^2 + \frac{1}{2} M^2_\Delta {\rm Tr} \left( \Delta^\dagger
\Delta \right) - \left[ \lambda^{}_\Delta M^{}_\Delta H^T i
\sigma^{}_2 \Delta H + {\rm h.c.} \right] \; .
\end{equation}
When the neutral components of $H$ and $\Delta$ acquire their vacuum
expectation values $v$ and $v^{}_\Delta$ respectively, the
electroweak gauge symmetry is spontaneously broken. The minimum of
$\cal V$ is achieved at $v = \mu/(\lambda -
2\lambda^2_\Delta)^{1/2}$ and $v^{}_\Delta = \lambda^{}_\Delta
v^2/M^{}_\Delta$, where $\lambda^{}_\Delta$ has been assumed to be
real. Note that $v^{}_\Delta$ may modify the SM masses of $W^\pm$
and $Z^0$ in such a way that $\rho \equiv M^2_W/(M^2_Z \cos^2
\theta^{}_{\rm W}) = (v^2 + 2v^2_\Delta)/(v^2 + 4v^2_\Delta)$ holds.
By using current experimental data on the
$\rho$-parameter,\cite{PDG} we get $\kappa \equiv \sqrt{2}
~v^{}_\Delta /v < 0.01$ and $v^{}_\Delta < 2.5~{\rm GeV}$. Given
$M^{}_\Delta \gg v$, an approximate seesaw formula turns out to be
\begin{equation}
M^{}_\nu \; \equiv \; M^{}_{\rm L} \; =\; Y^{}_\Delta v^{}_\Delta \;
\approx \; \lambda^{}_\Delta Y^{}_\Delta \frac{v^2}{M^{}_\Delta} \;
,
\end{equation}
as shown in Fig. 1. Note that the last term of Eq. (8) violates both
$L$ and $B-L$, and thus the smallness of $\lambda^{}_\Delta$ is
naturally allowed according to 't Hooft's naturalness criterion
(i.e., setting $\lambda^{}_\Delta =0$ will increase the symmetry of
the theory).\cite{Hooft} Given $M^{}_\Delta \sim {\cal O}(1)$ TeV,
for example, the seesaw works to generate $M^{}_\nu \sim {\cal
O}(10^{-2})$ eV provided $\lambda^{}_\Delta Y^{}_\Delta \sim {\cal
O}(10^{-12})$ holds. The neutrino mixing matrix $V$ is exactly
unitary in the type-II seesaw mechanism, simply because the heavy
degrees of freedom do not mix with the light ones.

There are totally seven physical Higgs bosons in this model:
doubly-charged $H^{++}$ and $H^{--}$, singly-charged $H^+$ and
$H^-$, neutral $A^0$ (CP-odd), and neutral $h^0$ and $H^0$
(CP-even), where $h^0$ is the SM-like Higgs boson. Except for
$M^2_{h^0} \approx 2\mu^2$, we get a quasi-degenerate mass spectrum
for other scalars: $M^2_{H^{\pm \pm}} = M^2_\Delta \approx
M^2_{H^0}$, $M^2_{H^\pm} = M^2_\Delta (1 + \kappa^2)$, and
$M^2_{A^0} = M^2_\Delta (1 + 2\kappa^2)$. As a consequence, the
decay channels $H^{\pm \pm} \to W^\pm H^\pm$ and $H^{\pm \pm} \to
H^\pm H^\pm$ are kinematically forbidden. The production of
$H^{\pm\pm}$ at the LHC is mainly through $q\bar{q} \to \gamma^*,
Z^* \to H^{++}H^{--}$ and $q\bar{q}^\prime \to W^* \to
H^{\pm\pm}H^\mp$ processes, which do not depend on the small Yukawa
couplings.

The typical collider signatures in this seesaw scenario are the
lepton-number-violating $H^{\pm\pm} \to l^\pm_\alpha l^\pm_\beta$
decays\cite{Han1} as well as $H^+ \to l^+_\alpha \overline{\nu}$ and
$H^- \to l^-_\alpha \nu$ decays.\cite{Han2} Their branching ratios
\begin{eqnarray}
{\cal B}(H^{\pm\pm} \to l^\pm_\alpha l^\pm_\beta) = \frac{( 2 -
\delta^{}_{\alpha\beta}) |(M^{}_{\rm
L})^{}_{\alpha\beta}|^2}{\displaystyle \sum_{\rho,\sigma}
|(M^{}_{\rm L})^{}_{\rho\sigma}|^2} \; , ~~~~ {\cal B}(H^+ \to
l^+_\alpha \overline{\nu}) = \frac{\displaystyle \sum_\beta
|(M^{}_{\rm L})^{}_{\alpha\beta}|^2}{\displaystyle
\sum_{\rho,\sigma} |(M^{}_{\rm L})^{}_{\rho\sigma}|^2} \; ~~~
\end{eqnarray}
are closely related to the masses, flavor mixing angles and
CP-violating phases of three light neutrinos, because $M^{}_{\rm
L} = V \widehat{M}^{}_\nu V^T$ holds. Some numerical analyses of
such decay modes together with the LHC signatures of $H^{\pm\pm}$
and $H^{\pm}$ bosons have been done by a number of authors (see,
e.g., Refs. \refcite{Han1} and \refcite{Han2}).

\subsection{Type-(I+II) seesaw}

A type-(I+II) seesaw mechanism can be achieved by combining the
neutrino mass terms in Eq. (1) with the Higgs potential in Eq. (8).
The seesaw formula is
\begin{equation}
M^{}_\nu \; \equiv \; V \widehat{M}^{}_\nu V^T \; \approx \;
M^{}_{\rm L} - M^{}_{\rm D} M^{-1}_{\rm R} M^T_{\rm D} \;
\end{equation}
in the leading-order approximation, where the small deviation of $V$
from unitarity has been omitted and the expression of $M^{}_{\rm L}$
can be found in Eq. (9). Hence the type-I and type-II seesaws can be
regarded as two extreme cases of the type-(I+II) seesaw. Note that
two mass terms in Eq. (11) are possibly comparable in magnitude. If
both of them are small, their contributions to $M^{}_\nu$ should
essentially be constructive; but if both of them are large, their
contributions to $M^{}_\nu$ must be destructive. The latter case
unnaturally requires a significant cancellation between two big
quantities in order to obtain a small quantity, but it is
interesting in the sense that it may give rise to observable
collider signatures of heavy Majorana neutrinos.\cite{Chao}

Let me briefly describe a type-(I+II) seesaw model and comment on
its possible LHC signatures.\cite{Si} First, we assume that both
$M^{}_i$ and $M^{}_\Delta$ are of ${\cal O}(1)$ TeV. Then the
production of $H^{\pm\pm}$ and $H^\pm$ bosons at the LHC is
guaranteed, and their lepton-number-violating signatures will probe
the Higgs triplet sector of the type-(I+II) seesaw
mechanism.\cite{Ren} On the other hand, ${\cal O}(M^{}_{\rm
D}/M^{}_{\rm R}) \lesssim {\cal O}(0.1)$ is possible as a result of
${\cal O}( M^{}_{\rm R}) \sim {\cal O}(1)$ TeV and ${\cal
O}(M^{}_{\rm D}) \lesssim {\cal O}(v)$, such that appreciable
signatures of $N^{}_i$ can be achieved at the LHC. Second, the small
mass scale of $M^{}_\nu$ implies that the relation ${\cal
O}(M^{}_{\rm L}) \sim {\cal O}(M^{}_{\rm D} M^{-1}_{\rm R} M^T_{\rm
D})$ must hold. In other words, it is the significant but incomplete
cancellation between $M^{}_{\rm L}$ and $M^{}_{\rm D} M^{-1}_{\rm R}
M^T_{\rm D}$ terms that results in the non-vanishing but tiny masses
for three light neutrinos. We admit that dangerous radiative
corrections to two mass terms of $M^{}_\nu$ require a delicate
fine-tuning of the afore-mentioned cancellation.\cite{Si} But this
scenario allows us to reconstruct $M^{}_{\rm L}$ via the excellent
approximation $M^{}_{\rm L} = V \widehat{M}^{}_\nu V^T + R
\widehat{M}^{}_N R^T \approx R \widehat{M}^{}_N R^T$, such that the
elements of the Yukawa coupling matrix $Y^{}_\Delta$ read
\begin{eqnarray}
\left(Y^{}_\Delta\right)^{}_{\alpha \beta} = \frac{\left(M^{}_{\rm
L}\right)^{}_{\alpha \beta}}{v^{}_\Delta} \approx \sum^3_{i=1}
\frac{R^{}_{\alpha i} R^{}_{\beta i} M^{}_i}{v^{}_\Delta} \; ,
\end{eqnarray}
where the subscripts $\alpha$ and $\beta$ run over $e$, $\mu$ and
$\tau$. This result implies that the leptonic decays of $H^{\pm
\pm}$ and $H^\pm$ bosons depend on both $R$ and $M^{}_i$, which
actually determine the production and decays of $N^{}_i$. Thus we
have established an interesting correlation between the singly- or
doubly-charged Higgs bosons and the heavy Majorana neutrinos. To
observe the correlative signatures of $H^\pm$, $H^{\pm\pm}$ and
$N^{}_i$ at the LHC will serve for a direct test of this type-(I+II)
seesaw model.

To illustrate, here I focus on the {\it minimal} type-(I+II) seesaw
model with a single heavy Majorana neutrino,\cite{Gu} where $R$ can
be parametrized in terms of three rotation angles $\theta^{}_{i4}$
and three phase angles $\delta^{}_{i4}$ (for
$i=1,2,3$).\cite{Xing08} In this case, we have
\begin{eqnarray}
\omega^{}_1 & \equiv & \frac{\sigma (pp \to \mu^+ \mu^+ W^-
X)|^{}_{N^{}_1}}{\sigma (pp \to \mu^+ \mu^+ H^- X)|^{}_{H^{++}}}
\; \approx \; \frac{\sigma^{}_N}{\sigma^{}_H} \cdot \frac{s^2_{14} +
s^2_{24} + s^2_{34}}{4} \; , \nonumber \\
\omega^{}_2 & \equiv & \frac{\sigma (pp \to \mu^+ \mu^+ W^-
X)|^{}_{N^{}_1}}{\sigma (pp \to \mu^+ \mu^+ H^{--} X)|^{}_{H^{++}}}
\; \approx \; \frac{\sigma^{}_N}{\sigma^{}_{\rm pair}} \cdot
\frac{s^2_{14} + s^2_{24} + s^2_{34}}{4} \;
\end{eqnarray}
for $s^{}_{i4} \equiv \sin\theta^{}_{i4} \lesssim {\cal O}(0.1)$,
where $\sigma^{}_{N} \equiv \sigma(pp \to l^+_\alpha N^{}_1
X)/|R^{}_{\alpha 1}|^2$, $\sigma^{}_{H} \equiv \sigma(pp \to H^{++}
H^- X)$ and $\sigma^{}_{\rm pair} \equiv \sigma(pp \to H^{++}
H^{--}X)$ are three reduced cross sections.\cite{Si} Fig. 2
illustrates the numerical results of $\omega^{}_1$ and $\omega^{}_2$
changing with $M^{}_1$ at the LHC with an integrated luminosity of
$300 ~ {\rm fb}^{-1}$, just to give one a ball-park feeling of
possible collider signatures of $N^{}_1$ and $H^{\pm\pm}$ and their
correlation in our model.
\begin{figure*}[t]
\centering
\includegraphics[width=90mm]{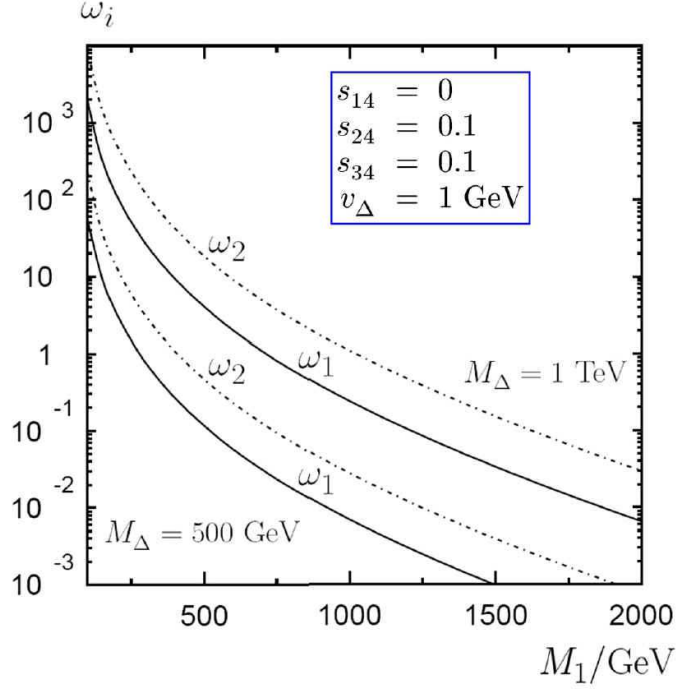}
\caption{Correlative signatures of $N^{}_1$ and $H^{\pm\pm}$ at
the LHC with a luminosity of $300~{\rm fb}^{-1}$.}
\end{figure*}

\section{Unitarity Violation}

It is worth emphasizing that the charged-current interactions of
light and heavy Majorana neutrinos are not completely independent
in either the type-I seesaw or the type-(I+II) seesaw. The
standard charged-current interactions of $\nu^{}_i$ and $N^{}_i$
are
\begin{eqnarray}
-{\cal L}^{}_{\rm cc} = \frac{g}{\sqrt{2}} \left[ \overline{\left(e~
\mu~ \tau\right)^{}_{\rm L}} ~V \gamma^\mu \left( \begin{matrix}
\nu^{}_1 \cr \nu^{}_2 \cr \nu^{}_3
\end{matrix} \right)^{}_{\rm L} W^-_{\mu} + \overline{\left(e~
\mu~ \tau\right)^{}_{\rm L}} ~R \gamma^\mu \left(
\begin{matrix} N^{}_1 \cr N^{}_2 \cr N^{}_3 \end{matrix}
\right)^{}_{\rm L} W^-_\mu \right] + {\rm h.c.} , ~~~
\end{eqnarray}
where $V$ is just the light neutrino mixing matrix responsible for
neutrino oscillations, and $R$ describes the strength of
charged-current interactions between $(e, \mu, \tau)$ and
$(N^{}_1, N^{}_2, N^{}_3)$. Since $V$ and $R$ belong to the same
unitary transformation done in Eq. (4), they must be correlated
with each other and their correlation signifies an important
relationship between neutrino physics and collider physics.

It has been shown that $V$ and $R$ share nine rotation angles
($\theta^{}_{i4}$, $\theta^{}_{i5}$ and $\theta^{}_{i6}$ for
$i=1$, $2$ and $3$) and nine phase angles ($\delta^{}_{i4}$,
$\delta^{}_{i5}$ and $\delta^{}_{i6}$ for $i=1$, $2$ and
$3$).\cite{Xing08} To see this point clearly, let me decompose $V$
into $V = A V^{}_0$, where
\begin{equation}
V^{}_0 = \left( \begin{matrix} c^{}_{12} c^{}_{13} & \hat{s}^*_{12}
c^{}_{13} & \hat{s}^*_{13} \cr -\hat{s}^{}_{12} c^{}_{23} -
c^{}_{12} \hat{s}^{}_{13} \hat{s}^*_{23} & c^{}_{12} c^{}_{23} -
\hat{s}^*_{12} \hat{s}^{}_{13} \hat{s}^*_{23} & c^{}_{13}
\hat{s}^*_{23} \cr \hat{s}^{}_{12} \hat{s}^{}_{23} - c^{}_{12}
\hat{s}^{}_{13} c^{}_{23} & -c^{}_{12} \hat{s}^{}_{23} -
\hat{s}^*_{12} \hat{s}^{}_{13} c^{}_{23} & c^{}_{13} c^{}_{23} \cr
\end{matrix} \right)
\end{equation}
with $c^{}_{ij} \equiv \cos\theta^{}_{ij}$ and $\hat{s}^{}_{ij}
\equiv e^{i\delta^{}_{ij}} \sin\theta^{}_{ij}$ is just the
standard parametrization of the $3\times 3$ unitary neutrino
mixing matrix (up to some proper phase
rearrangements).\cite{PDG,Xing04} Because of $VV^\dagger =
AA^\dagger = {\bf 1} - RR^\dagger$, it is obvious that $V
\rightarrow V^{}_0$ in the limit of $A \rightarrow {\bf 1}$ (or
equivalently, $R \rightarrow {\bf 0}$). Considering the fact that
the non-unitarity of $V$ must be a small effect (at most at the
percent level as constrained by current neutrino oscillation data
and precision electroweak data\cite{Antusch}), we expect
$s^{}_{ij} \lesssim {\cal O}(0.1)$ (for $i=1,2,3$ and $j=4,5,6$)
to hold. Then we obtain\cite{Xing08}
\begin{eqnarray}
A & = & {\bf 1} - \left( \begin{matrix} \frac{1}{2} \left( s^2_{14}
+ s^2_{15} + s^2_{16} \right) & 0 & 0 \cr \hat{s}^{}_{14}
\hat{s}^*_{24} + \hat{s}^{}_{15} \hat{s}^*_{25} + \hat{s}^{}_{16}
\hat{s}^*_{26} & \frac{1}{2} \left( s^2_{24} + s^2_{25} + s^2_{26}
\right) & 0 \cr \hat{s}^{}_{14} \hat{s}^*_{34} + \hat{s}^{}_{15}
\hat{s}^*_{35} + \hat{s}^{}_{16} \hat{s}^*_{36} & \hat{s}^{}_{24}
\hat{s}^*_{34} + \hat{s}^{}_{25} \hat{s}^*_{35} + \hat{s}^{}_{26}
\hat{s}^*_{36} & \frac{1}{2} \left( s^2_{34} +
s^2_{35} + s^2_{36} \right) \cr \end{matrix} \right) , \nonumber \\
R & = & {\bf 0} + \left( \begin{matrix} \hat{s}^*_{14} &
\hat{s}^*_{15} & \hat{s}^*_{16} \cr \hat{s}^*_{24} & \hat{s}^*_{25}
& \hat{s}^*_{26} \cr \hat{s}^*_{34} & \hat{s}^*_{35} &
\hat{s}^*_{36} \cr \end{matrix} \right)
\end{eqnarray}
as two excellent approximations. A striking consequence of the
non-unitarity of $V$ is the loss of universality for the Jarlskog
invariants of CP violation,\cite{J} $J^{ij}_{\alpha\beta} \equiv
{\rm Im}(V^{}_{\alpha i} V^{}_{\beta j} V^*_{\alpha j} V^*_{\beta
i})$, where the Greek indices run over $(e, \mu, \tau)$ and the
Latin indices run over $(1,2,3$). For example, the extra
CP-violating phases of $V$ are possible to give rise to a
significant asymmetry between $\nu^{}_\mu \rightarrow \nu^{}_\tau$
and $\overline{\nu}^{}_\mu \rightarrow \overline{\nu}^{}_\tau$
oscillations.

The probability of $\nu^{}_\alpha \rightarrow \nu^{}_\beta$
oscillations in vacuum, defined as $P^{}_{\alpha\beta}$, is given
by\cite{Xing08}
\begin{eqnarray}
P^{}_{\alpha\beta} \; = \frac{\displaystyle \sum^{}_i
|V^{}_{\alpha i}|^2 |V^{}_{\beta i}|^2 + 2 \sum^{}_{i<j} {\rm Re}
\left( V^{}_{\alpha i} V^{}_{\beta j} V^*_{\alpha j} V^*_{\beta i}
\right) \cos \Delta^{}_{ij} - 2 \sum^{}_{i<j} J^{ij}_{\alpha\beta}
\sin\Delta^{}_{ij}}{\displaystyle \left(
VV^\dagger\right)^{}_{\alpha\alpha} \left(
VV^\dagger\right)^{}_{\beta\beta}} \; , ~~~
\end{eqnarray}
where $\Delta^{}_{ij} \equiv \Delta m^2_{ij} L/(2E)$ with $\Delta
m^2_{ij} \equiv m^2_i - m^2_j$, $E$ being the neutrino beam energy
and $L$ being the baseline length. If $V$ is exactly unitary
(i.e., $A = {\bf 1}$ and $V = V^{}_0$), the denominator of Eq.
(17) will become unity and the conventional formula of
$P^{}_{\alpha\beta}$ will be reproduced. It has been observed in
Refs. \refcite{Xing08} and \refcite{Yasuda} that $\nu^{}_\mu
\rightarrow \nu^{}_\tau$ and $\overline{\nu}^{}_\mu \rightarrow
\overline{\nu}^{}_\tau$ oscillations may serve as a good tool to
probe possible signatures of CP violation induced by the
non-unitarity of $V$. To illustrate this point, we consider a
short- or medium-baseline neutrino oscillation experiment with
$|\sin\Delta^{}_{13}| \sim |\sin\Delta^{}_{23}| \gg
|\sin\Delta^{}_{12}|$, in which the terrestrial matter effects are
expected to be insignificant or negligibly small. Then the
dominant CP-conserving and CP-violating terms of $P(\nu^{}_\mu
\rightarrow \nu^{}_\tau)$ and $P(\overline{\nu}^{}_\mu \rightarrow
\overline{\nu}^{}_\tau)$ can simply be obtained from Eq. (17):
\begin{eqnarray}
P(\nu^{}_\mu \rightarrow \nu^{}_\tau) & \approx & \sin^2
2\theta^{}_{23} \sin^2 \frac{\Delta^{}_{23}}{2} ~ - ~ 2 \left(
J^{23}_{\mu\tau} + J^{13}_{\mu\tau} \right) \sin\Delta^{}_{23} \;
, \nonumber \\
P(\overline{\nu}^{}_\mu \rightarrow \overline{\nu}^{}_\tau) &
\approx & \sin^2 2\theta^{}_{23} \sin^2 \frac{\Delta^{}_{23}}{2} ~
+ ~ 2 \left( J^{23}_{\mu\tau} + J^{13}_{\mu\tau} \right)
\sin\Delta^{}_{23} \; ,
\end{eqnarray}
where the good approximation $\Delta^{}_{13} \approx
\Delta^{}_{23}$ has been used in view of the experimental fact
$|\Delta m^2_{13}| \approx |\Delta m^2_{23}| \gg |\Delta
m^2_{12}|$, and the sub-leading and CP-conserving ``zero-distance"
effect\cite{Antusch} has been omitted. For simplicity, I take
$V^{}_0$ to be the exactly tri-bimaximal mixing pattern\cite{TB}
(i.e., $\theta^{}_{12} = \arctan(1/\sqrt{2})$, $\theta^{}_{13} =0$
and $\theta^{}_{23} =\pi/4$ as well as $\delta^{}_{12} =
\delta^{}_{13} = \delta^{}_{23} =0$) and then arrive
at\cite{Xing08}
\begin{eqnarray}
2\left( J^{23}_{\mu\tau} + J^{13}_{\mu\tau} \right) \; \approx \;
\sum^6_{l=4} s^{}_{2l} s^{}_{3l} \sin \left( \delta^{}_{2l} -
\delta^{}_{3l} \right) \; .
\end{eqnarray}
Given $s^{}_{2l} \sim s^{}_{3l} \sim {\cal O}(0.1)$ and
$(\delta^{}_{2l} - \delta^{}_{3l}) \sim {\cal O}(1)$ (for
$l=4,5,6$), this non-trivial CP-violating quantity can reach the
percent level. A numerical illustration of the CP-violating
asymmetry between $\nu^{}_\mu \rightarrow \nu^{}_\tau$ and
$\overline{\nu}^{}_\mu \rightarrow \overline{\nu}^{}_\tau$
oscillations has been presented in Ref. \refcite{Xing08}, from
which one can see that it is possible to measure this asymmetry in
the range $L/E \sim (100 \cdots 400)$ km/GeV if the experimental
sensitivity is $\leq 1\%$. A neutrino factory with the beam energy
$E$ being above $m^{}_\tau \approx 1.78$ GeV may have a good
chance to explore the non-unitary effect of CP
violation.\cite{Yasuda}

When a long-baseline neutrino oscillation experiment is concerned,
however, the terrestrial matter effects must be taken into account
because they might fake the genuine CP-violating
signals.\cite{Xing00} As for $\nu^{}_\mu \rightarrow \nu^{}_\tau$
and $\overline{\nu}^{}_\mu \rightarrow \overline{\nu}^{}_\tau$
oscillations discussed above, the dominant matter effect results
from the neutral-current interactions\cite{Luo} and modifies the
CP-violating quantity of Eq. (18) in the following way:
\begin{eqnarray}
2\left( J^{23}_{\mu\tau} + J^{13}_{\mu\tau} \right) \;
\Longrightarrow \; \sum^6_{l=4} s^{}_{2l} s^{}_{3l} \left[ \sin
\left( \delta^{}_{2l} - \delta^{}_{3l} \right) + A^{}_{\rm NC} L
\cos \left( \delta^{}_{2l} - \delta^{}_{3l} \right) \right] \; ,
\end{eqnarray}
where $A^{}_{\rm NC} = G^{}_{\rm F} N^{}_n /\sqrt{2}~$ with
$N^{}_n$ being the background density of neutrons, and $L$ is the
baseline length. It is easy to find $A^{}_{\rm NC} L \sim {\cal
O}(1)$ for $L \sim 4 \times 10^3$ km.

\section{Concluding Remarks}

We hope that the LHC might open a new window for us to understand
the origin of neutrino masses and the dynamics of lepton number
violation. To be more specific, a TeV seesaw might work ({\it
naturalness}?) and its heavy degrees of freedom might show up at
the LHC ({\it testability}?). A bridge between collider physics
and neutrino physics is highly anticipated and, if it exists, will
lead to rich phenomenology.

\section*{Acknowledgments}

I am greatly indebted to A.H. Chan, C.H. Oh and K.K. Phua for warm
hospitality during my visiting stay in Singapore. I am also
grateful to W. Chao, Z. Si and S. Zhou for enjoyable
collaboration. This work is supported in part by the National
Natural Science Foundation of China under grant No. 10425522 and
No. 10875131.

\end{document}